\documentclass[preprint,12pt,authoryear]{elsarticle}

\usepackage{amssymb}
\usepackage{amsmath}
\usepackage{color}
\usepackage{multirow}
\usepackage{subfigure}

\journal{IEEE Transactions on Industrial Informatics}

\begin{document}
	
	\begin{frontmatter}
		
		
		
		\title{Single-Shared Network with Prior-Inspired Loss for Parameter-Efficient Multi-Modal Imaging Skin Lesion Classification}

		\author[label1,label2]{Peng TANG}
		\author[label1,label2]{Tobias Lasser}

		\address[label1]{organization={Department of Informatics, School of Computation, Information, and Technology, Technical University of Munich},
			city={Garching},
			country={Germany}}
		
		\address[label2]{organization={Munich Institute of Biomedical Engineering, Technical University of Munich},
			city={Garching},
			country={Germany}
		}

		\begin{abstract}
		In this study, we introduce a multi-modal approach that efficiently integrates multi-scale clinical and dermoscopy features within a single network, thereby substantially reducing model parameters. The proposed method includes three novel fusion schemes.
		Firstly, unlike current methods that usually employ two individual models for for clinical and dermoscopy modalities, we verified that multimodal feature can be learned by sharing the parameters of encoder while leaving the individual modal-specific classifiers.
		Secondly, the shared cross-attention module can replace the individual one to efficiently interact between two modalities at multiple layers.
		Thirdly, different from current methods that equally optimize dermoscopy and clinical branches, inspired by prior knowledge that dermoscopy images play a more significant role than clinical images, we propose a novel biased loss. This loss guides the single-shared network to prioritize dermoscopy information over clinical information, implicitly learning a better joint feature representation for the modal-specific task.
		Extensive experiments on a well-recognized Seven-Point Checklist (SPC) dataset and a collected dataset demonstrate the effectiveness of our method on both CNN and Transformer structures. Furthermore, our method exhibits superiority in both accuracy and model parameters compared to currently advanced methods.
		\end{abstract}
		
		
		
		
	\end{frontmatter}
	
	
	
	\section{Introduction}
	
	The skin serves as the body's largest organ, safeguarding against external threats and invasion. Additionally, it plays crucial roles in thermoregulation, metabolism and sensory perception the body \citep{kolarsick2011anatomy}.
	Over the past 30 years, skin cancer has emerged as one of the most lethal and rapidly spreading malignancies worldwide \citep{siegel2022cancer}.
	Among skin cancers, melanoma is the most fatal, as it can rapidly metastasize throughout the body and lead to a painful death.
	The five-year survival rate of melanoma can be improved to 95$\%$ if it is treated at an early stage \citep{balch2009final}. However, early prevention of melanoma is hindered by a large number of patients with skin diseases and a shortage of experienced dermatologists
	Therefore, there is an expectation that utilizing deep learning (DL) methods as an automatic aided system can enhance the diagnostic accuracy and efficiency of dermatologists.
	
	With the development of deep learning, single-modality-based methods have experienced significant improvements compared to former hand-crafted methods. However, from a data-driven perspective, deep learning models tend to achieve more accurate predictions when they are provided with more information. Therefore, an increasing number of researchers have begun to explore the complementary information between clinical and dermoscopy images to achieve more robust results in complex clinical scenarios.
	\citep{kawahara2018seven, yap2018multimodal} were among the first to propose fusing multi-modal features using concatenation for skin lesion classification.
	Subsequent research of \citep{tang2022fusionm4net, fu2022graph} improved performance by integrating prediction information in addition to feature fusion.
	To further enhance the diagnostic accuracy, \citep{bi2020multi, he2023co, zhang2023tformer} introduced more advanced fusion modules for the feature interaction of clinical and dermoscopy images. They argued that simple concatenation cannot fully exploit the information from both modalities.
	However, the introduction of fusion modules requires significant computational costs, limiting their applications in real-world scenarios.
	
	\begin{figure*}
	\centering
	\includegraphics[height=8cm,width=14cm]{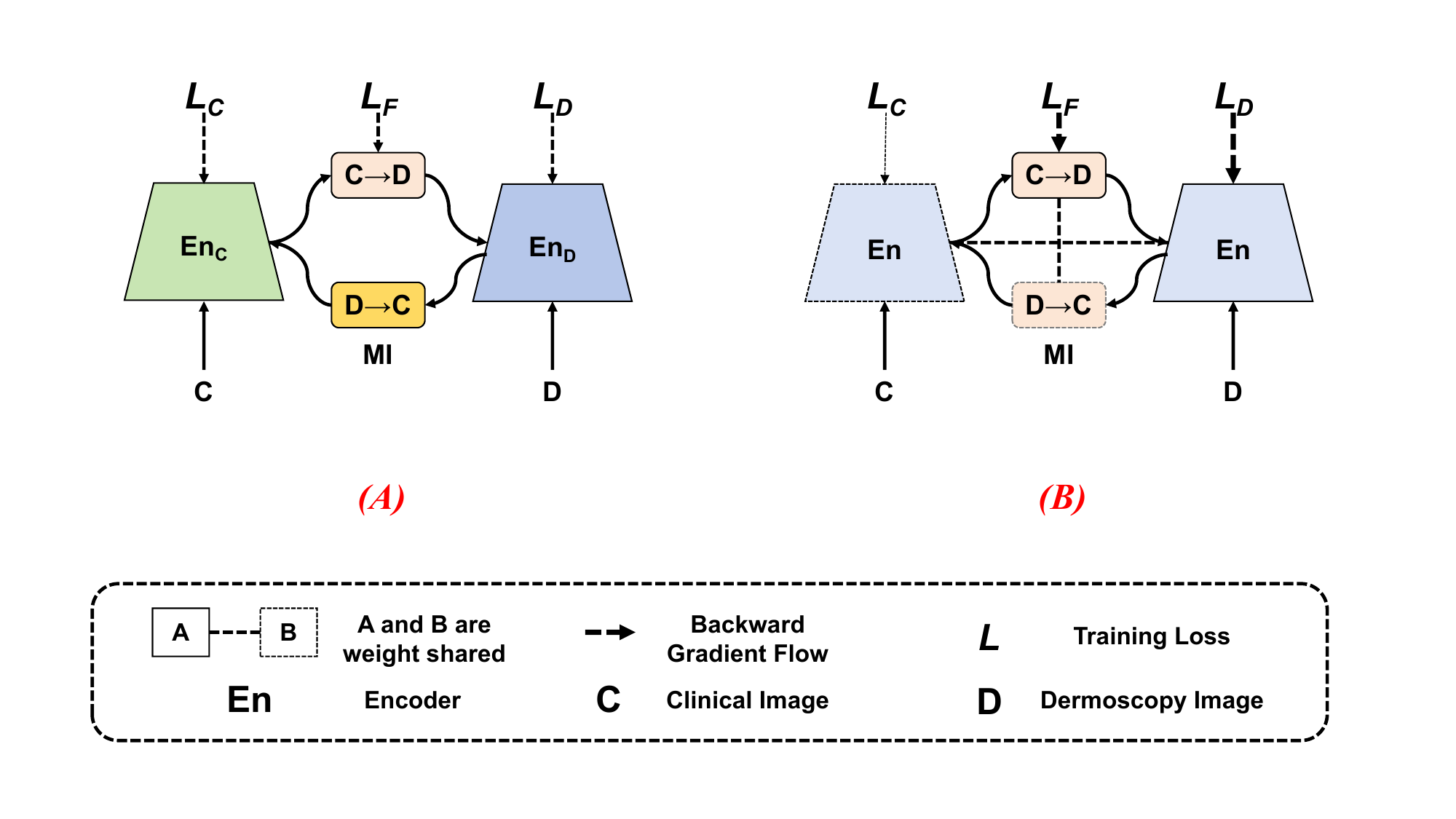}
	\vspace{-0.35in}	
	\caption{The overview structure of former methods and our PEMM framework.}
	
	\vspace{-0.1in}
	\label{fig1}
	
	\end{figure*}		
	In this paper, we propose a novel parameter-efficient multi-modal (PEMM) framework for skin lesion classification, achieving state-of-the-art classification performance while using fewer parameters compared to current advanced methods. 
	There are four difference between previous methods and our method.
	Firstly, unlike previous approaches that commonly employed ResNet as feature encoder, we conduct a comprehensive comparison between ResNet and more advanced backbones, i.e.,  DenseNet \citep{huang2017densely}, ConvNext (CXT) \citep{liu2022convnet}, and SwinTransformer (ST) \citep{liu2021swin}, which demonstrate that the latter three backbones can achieve higher accuracy with fewer parameters compared to ResNet.
	Secondly, given that the encoder accounts for the majority of the model's parameters, we naturally consider the idea of fusing multimodal features within a single network rather than using two individual encoders (See Fig.~\ref{fig1}). 
	Therefore, we explore and verify that multimodal features can be efficiently learned in a single-shared network with strong capacity by merely remaining the modal-specific classifiers, such as CXT and ST, resulting in significant parameter reduction while maintaining or subtly affecting accuracy.
	Thirdly, building on the concept of a 'shared network', we extend it to the fusion module and introduce a new shared cross-attention mechanism to efficiently conduct modality interaction on multi-scale multimodal features.
	Finally, inspired by the prior knowledge that dermoscopy images provide more useful information for diagnosis than clinical images, we introduce a new biased loss function. This function enables the model to focus more on the dermoscopy branch and less on the clinical branch, learning a better joint feature representation for the modal-specific classification task.
	Evaluations were conducted on two public datasets, and the results demonstrate the superiority of the proposed PEMM framework in both accuracy and model parameter efficiency compared to current state-of-the-art methods. Extensive experiments validate the effectiveness of our method across both CNN and Transformer structures.
	The main contributions of our method can be summarized as follows:
	\begin{enumerate}

	\item We validated that both clinical and dermoscopy modalities can be input into a single-shared network with strong capacity, achieving similar performance while reducing a large number of parameters compared to commonly-used two individual networks.
	
	\item We introduced a new shared cross-attention module to efficiently integrate multimodal features at different layers.
	
	\item We propose a novel prior-biased loss that guides the single-shared network to learn more meaningful information for accurate diagnosis.
	
	\item Our fusion method significantly outperforms state-of-the-art fusion methods, with only about 1 additional parameter increase on single-modal-based networks.
	
	\end{enumerate}
	
	\section{related works}
	\subsection{single-modal imaging based methods for skin lesion classification}
	Most current deep learning-based skin lesion classification methods predict results based on single-modal images, such as clinical or dermoscopy images. Clinical images, captured by standard digital cameras or smartphone cameras \citep{ge2017skin, pacheco2020pad}, primarily display the geometry and color of the lesion \citep{yang2018clinical}. 
	
	To facilitate research in this area, \citep{sun2016benchmark} released a large clinical skin diseases dataset, known as SD-198, serving as a benchmark for comparison between convolutional neural networks (CNNs) and hand-crafted features.
	\citep{yang2018clinical} presented effective feature representations by incorporating dermatologist's criteria, enhancing diagnostic performance, and capturing the manifestations of skin lesions. 
	Additionally, they introduced a new metric called the 'complexity of image category' to guide self-paced balanced learning, addressing the class-imbalanced problem in classification tasks \citep{yang2019self}. 
	Compared to hand-crafted methods, DL-based methods have achieved significant improvements in clinical-image (CI) based skin lesion classification. However, there still exists a considerable gap between CI-based and dermoscopy image-based methods \citep{togawa2023comparison, dascalu2022non}. For instance, in a comparative study by \citep{dascalu2022non}, dermoscopy image (DI)-based CNN models significantly increased the accuracy of skin cancer diagnosis from 75$\%$ to 88$\%$ when compared to smartphone images
	
	More researches \citep{esteva2017dermatologist, gu2019progressive, tang2020gp, yao2021single, liu2022ci, gao2024multi} is directed towards dermoscopy images (DI) rather than clinical images (CI) due to two main factors. Firstly, as mentioned earlier, dermoscopy images offer higher diagnostic accuracy compared to clinical images \citep{dascalu2022non}. Secondly, the availability and high quality of numerous dermoscopy image datasets in challenges organized by the International Skin Imaging Collaboration also plays a significant role.
	\citep{esteva2017dermatologist} demonstrated that a CNN trained using 129,450 images can achieve comparable performance with 21 board-certificated dermatologists on both classification tasks: benign seborrheic keratoses vs. keratinocyte carcinomas and benign nevi vs. malignant melanomas.
	\citep{gu2019progressive} presented a progressive transfer learning method to address the generalization ability problem of fully-supervised methods and improve recognition performance, where adversarial learning was introduced to learn invariant attributes.
	\citep{yao2021single} combined several techniques, including DropOut-related regularization, modified RandAugment, and a multi-weighted new loss, to address the class-imbalanced problem of skin lesion datasets.
	\citep{gao2024multi} explored and integrated information from different views, including RGB, HSL, and YCbCr, rather than only the RGB view, thereby enhancing skin lesion classification.	

	\subsection{multi-modal imaging based methods for skin lesion classification}
	Despite their success, current methods are developed based on only a single modality, thus not exploiting the complementary information provided by both clinical and dermoscopy images \citep{he2023co}. There is a growing body of research focused on utilizing multi-modal images for skin lesion classification \citep{ge2017skin, kawahara2018seven, yap2018multimodal, bi2020multi, tang2022fusionm4net, wang2022adversarial, fu2022graph, he2023co, zhang2023tformer}.
	\citep{kawahara2018seven} released the first multi-modal dataset for multi-label skin lesion classification, known as the Seven-point Checklist (SPC) dataset, laying the foundation stone in this field.
	\citep{tang2022fusionm4net, fu2022graph} improved upon previous methods by incorporating a late fusion scheme to combine both feature and prediction information from different modalities. \citep{tang2022fusionm4net} employed a weighted averaging scheme, while \citep{fu2022graph} adopted a graph learning technique.
	\citep{wang2022adversarial} proposed a modality discriminator that guides feature encoders to capture correlated and complementary information from two modalities.
	The most advanced methods \citep{bi2020multi, he2023co, zhang2023tformer} focused on designing an additional fusion branch to facilitate modality interaction using both extracted features at multiple layers. The branch in HcCNN \citep{bi2020multi} was based on concatenation-based fusion and multi-scale attention modules, while CFANet \citep{he2023co} and TFormer \citep{zhang2023tformer} mainly utilized cross-attention modules.
	However, the introduction of additional fusion branches incurs significant computational costs, which may hinder their application in various scenarios, such as deploying on mobile devices or implementing local AI-enabled family doctor systems for skincare.

	\subsection{Parameter-Sharing Network}
	Parameter-sharing network (PSN) or weight-sharing networks (WSN) are commonly employed in self-supervised learning as siamese networks. They are fed with multiple variants from the same source and then minimize the loss between their corresponding outputs to obtain task-related feature representations \citep{huang2022learning, schurholt2021self, tao2022exploring}. Additionally, some works utilize PSN to improve performance while achieving lower memory consumption \citep{aich2020multi, wang2023adaptive, wang2020learning}.
	For instance, \citep{wang2023adaptive} presented a parameter-sharing transformer block that captures scale-invariant information for 3D medical image segmentation. Similarly, \citep{wang2020learning} introduced a WSN that efficiently fuses RGB images and depth input for semantic segmentation tasks.
	However, there is a significant gap between the application scenarios due to the different types of data and tasks. Therefore, these methods cannot be directly applied to our task.
	
	In the task of multi-modal skin lesion classification, TFormer \citep{zhang2023tformer} employed a weight-sharing scheme to alleviate the overfitting problem. However, they did not thoroughly explore the impact of weight-sharing schemes on reducing parameters, leading to a confusing conclusion.
	For instance, in their configuration, the parameters of the introduced fusion branch are nearly identical to those of the feature encoder. It is highly probable that the weight-sharing scheme is achieved through the fusion branch rather than the encoder's capacity.
	In this paper, we verified that the single-shared network for parameter reduction is achieved based on encoder's capacity and maintaining individual classifiers, and further explored its generalization ability across different backbones by conducting extensive experiments.
	Moreover, in comparison to TFormer, we propose a new shared cross-attention module to efficiently reduce parameters on the fusion branch. Additionally, we introduce a novel biased loss mechanism that guides the single-shared network to be better optimized for the classification task.

	\section{Method: Parameter-Efficient Multi-Modal (PEMM) framework}
\begin{figure*}[h]
	\centering
	\includegraphics[height=8cm,width=14cm]{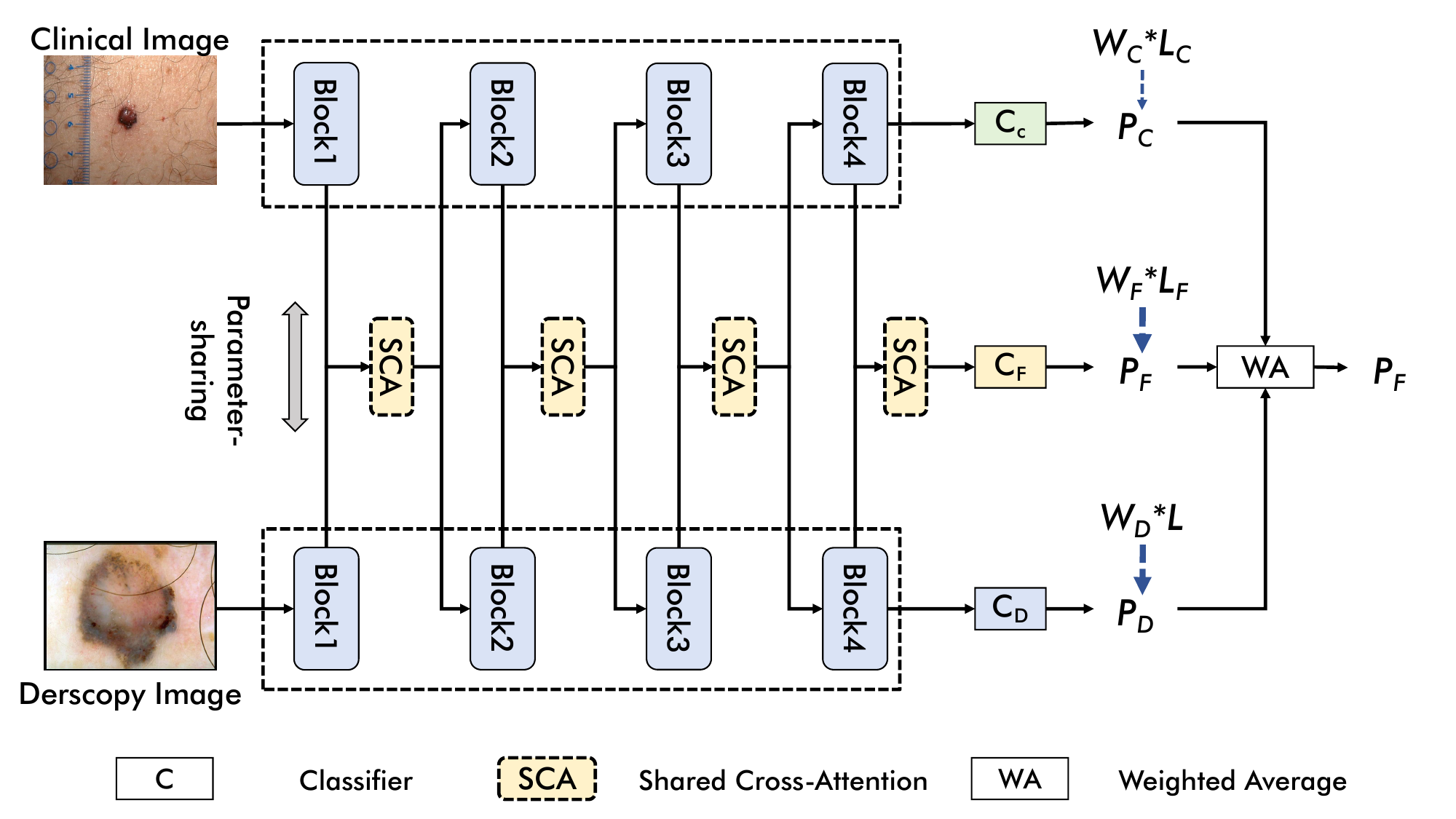}
	\caption{The detailed pipeline of our PEMM framework.}
	\vspace{-0.1in}
	\label{fig2}
	\end{figure*}

	The first step of our work is to explore utilizing different backbones as feature encoders instead of directly using ResNet for our classification task. Since many advanced backbones have been proposed and achieved better performance than ResNet for natural image recognition, such as DenseNet, ConvNext, and SwinTransformer. 
	The results in Table \ref{tab9} demonstrate the superiority of advanced backbones in improving classification accuracy and parameter reduction compared to the commonly-used ResNet.
	After that, we gradually introduce three main components, namely a single-shared network, shared cross-attention modules, and a biased loss function, as shown in Figure \ref{fig2}, into our Parameter-Efficient Multi-Modal (PEMM) framework.
	
	\subsection{Single-Shared Network}
	Following \citep{tang2022fusionm4net, fu2022graph, he2023co}, we also adopt two extra classifiers that can predict on clinical $C_C$ and dermoscopy $C_D$ branches  and then conduct the late fusion on the prediction-level for more accurate results. Therefore, our baseline model contains two individual encoders, and three individual classifiers.
	
	In deep learning-based multi-modal methods, feature encoders are indispensable as they are responsible for extracting individual features from different modalities, often occupying the majority of parameters in the entire model.
	Therefore, to build a parameter-efficient multi-modal method, we explore the extraction of modality-specific features from both clinical and dermoscopy images using a single-shared encoder (as illustrated in Fig.\ref{fig1}(b)), rather than using two individual encoders as commonly done in previous methods (See Fig.\ref{fig1}(a)).
	More specifically, as depicted in Fig.~\ref{fig2}, the Single-Shared Network (SSN) adopts weight-sharing encoders to extract multi-modal features, while individual classifiers are built upon fully connected layers to predict on the extracted modality-specific features.
	We also attempted to share the parameters of the classifiers of dermoscopy and clinical branches, denoted as $C_D$ and $C_C$ respectively. However, the results were unsatisfactory, which is attributed to the robustness of convolution layers and the sensitivity of fully connected layers (More details can be found in Table~\ref{tab8}). 
	While this parameter-sharing scheme significantly compresses the parameters of our multi-modal fusion model, it is only effective with the ConvNext and SwinTransformer backbones. It fails to maintain accuracy compared to the corresponding non-parameter-sharing fusion model when ResNet and DenseNet are used as encoders (See Table)

	\begin{figure*}[h]
	\centering
	\includegraphics[height=8cm,width=14cm]{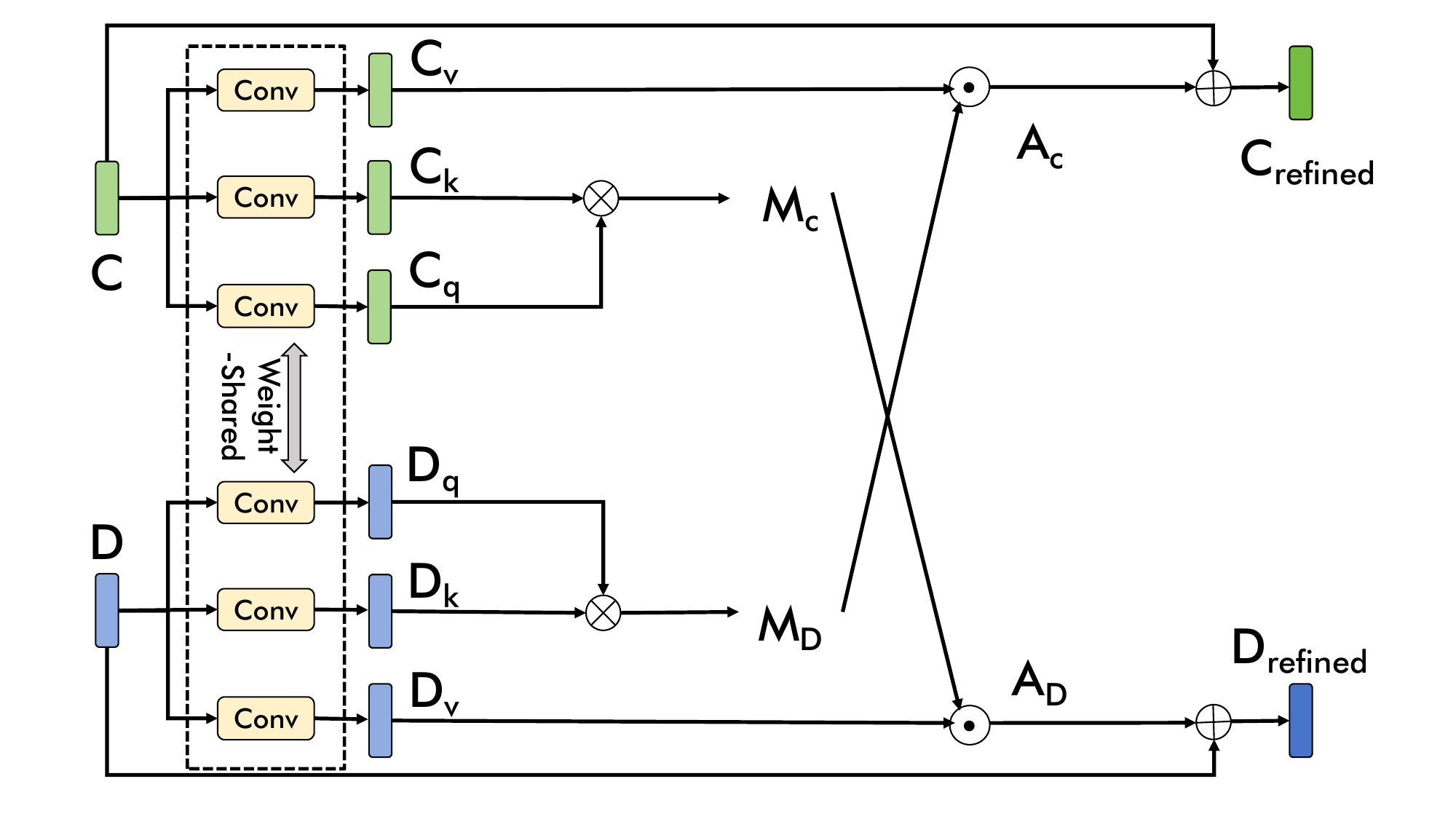}
	\caption{The detailed pipeline of shared cross-attention module.}
	\vspace{-0.1in}
	\label{fig3}
	\end{figure*}		
	
	\subsection{Shared Cross-Attention Module}
	The effectiveness of the current advanced fusion module, i.e., cross-attention (CA) for multi-modal skin image fusion has been demonstrated in \citep{he2023co}. As illustrated in Fig.~\ref{fig3}, the CA module employs three individual convolutions to project the input clinical feature $C$ into three feature vectors: $C_k$, $C_v$, and $C_q$. Subsequently, $C_v$ and $C_q$ are utilized to generate the attention map $M_c$ through feature transformation and matrix multiplication. A dot product operation is then applied to $C_v$ and $M_c$ to obtain the attentive features $A_c$. Finally, the refined clinical feature $C_{refined}$ is obtained through matrix summation between the input feature $C$ and $A_c$. Additional three convolutions are necessary to refine the dermoscopy feature $D_{refined}$.
	
	Following the concept of "parameter-sharing", we further refined the CA modules by sharing the parameters of the three convolutions for the projections of input features from both modalities (refer to Fig.~\ref{fig3}). Consequently, we can save half of the parameters of each CA module.

	\subsection{Biased Loss Function}
	In the training of previous methods, three branches are equally optimized, so their loss function can be formulated as Eq.~\ref{eq1}:
	\begin{equation}
	L_{total} = (L_C + L_D + L_F)/3  
	\label{eq1}
	\end{equation}	
	where $L_{total}$ represents the total loss function and $L_C$, $L_D$ and $L_F$ are the loss function for clinical, dermoscopy and fusion branches, respectively (See Fig.~\ref{fig3}).
	
	However, equally optimization of these three branches seems like not reasonable based on the prior knowledge, which demonstrated that  dermoscopy image-based model outperforms clinical image-based model \citep{dascalu2022non}.
	Inspired by the prior knowledge, we can have a hypothesis that dermoscopy information is more useful than clinical one in the multi-modal task and thus proposed a new biased loss function, which are achieved by adjusting the corresponding weights of loss functions for different branches, as shwon in Eq.~\ref{eq2}. 
	\begin{equation}
		L_{total} = W_C \cdot L_C + W_D \cdot L_D + W_F \cdot L_F  
		\label{eq2}
	\end{equation}	
	where $W_C$, $W_D$ and $W_F$ are the corresponding weights of $L_C$, $L_D$ and $L_F$, respectively.
	Specifically, in this function, $W_D$ is set to bigger than $W_C$, and $W_F$ are the sum of  $W_C$ and $W_D$ as the fusion information is the combination of clinical and dermoscopy information. So,  Eq.~\ref{eq2} can be simplified into Eq.~\ref{eq3}
	\begin{equation}
		L_{total} = W \cdot L_C + (0.5-W) \cdot L_D + 0.5  \cdot L_F  
		\label{eq3}
	\end{equation}	
	Where $W$ is the weight factor and $W \in [0, 0.1, 0.2, 0.3, 0.4]$.
	With using this loss function, more backward gradient flows will pass the dermoscopy and branches and explicitly enforce the multi-modal model to concentrate more on the information from these two branches than clinical branch. 
	
	\section{Experiments}
	\subsection{Implementation Details}
	During training, the Adam optimizer \citep{kingma2014adam} is employed with a batch size of 24.
	The initial learning rate is set to 3e-5 and is adjusted every epoch following the CosineAnnealing learning schedule.
	Random transformations such as vertical and horizontal flipping, rotation, shifting, and enhancing brightness and contrast are applied during training.
	Stochastic weight averaging \citep{izmailov2018averaging} is utilized to generate the final weight used for testing.
	We set varying training epochs of 250 and 150 to the SPC dataset and the collected ISIC dataset, respectively, owing to disparities in data number and the complexity of classification tasks.
	All images are resized to 224 $\times$ 224 $\times$ 3 for both training and testing.
	During the testing, we followed \citep{tang2022fusionm4net} that searches the weights on the validation set and then form the final predictions by a weighted averaging scheme.
	All the experiments are based on the backbone of SwinTransformer are on the SPC dataset, unless specified.
	The weight factor in Eq.~\ref{eq3} is set to 0.1, as it yields the best performance (See Table~\ref{tab7}).
		
	\subsection{Datasets and Metrics}
	Two datasets are used to evaluate the effectiveness of our method: the Seven-Point Checklist (SPC) \citep{kawahara2018seven} and the collected ISIC dataset.
	
	\textbf{SPC dataset} The SPC dataset consists of 1011 cases, with each case containing both a dermoscopy image and a clinical image, along with diagnosis labels and seven-point checklist labels. For more detailed label information, refer to \citep{kawahara2018seven}. The dataset was pre-split by the creator, so we followed the default setting in our experiments.
	In the task of multi-label skin lesion classification, we follow previous methods \citep{kawahara2018seven, tang2022fusionm4net, he2023co, fu2022graph} by using the area under the curve (Avg AUC) and accuracy (Acc) as comparison metrics. Additionally, we provide precision (Prec), specificity, and sensitivity for supplementary analysis.
	
	\textbf{Collected ISIC dataset} We collected 290 pairs of clinical and dermoscopy images from the  ISIC Archive to build an additional multi-modal skin image dataset, named the collected ISIC dataset. It consists of 109 benign and 181 malignant cases.
	The collected ISIC dataset is divided into training, validation, and testing sets based on commonly used ratios of 0.7, 0.1, and 0.2, respectively.
	For evaluation, we used the metrics of averaged precision (AP) from the ISIC 2016 skin lesion classification challenge and Avg AUC. These metrics are chosen because the challenge involves classifying lesions into benign and malignant categories, aligning well with our task.

\subsection{Comparison with state-of-the-art methods}
We undertake a comparative analysis of our PEMM model with several existing methodologies, including TFormer \citep{zhang2023tformer}, GIIN \citep{fu2022graph}, FusionM4Net-FS \citep{tang2022fusionm4net}, AMFAM \citep{wang2022adversarial}, HcCNN \citep{bi2020multi}, and Inception-combination \citep{kawahara2018seven}, on the SPC dataset. The comparative results concerning Averaged AUC and accuracy are presented in Tables~\ref{tab1} and \ref{tab2}, respectively.
It is noteworthy that all reported results are extracted from the respective literature and are presumed to represent the optimal performance of each model, except for TFormer, which reported an averaged accuracy value. Therefore, for the comparison, we also opt for the model's weights demonstrating the best performance in terms of Avg AUC. For the ensuing experiments, our model underwent training five times, and the mean values alongside the standard deviation from these five iterations were employed for a more robust analysis of our model.

As demonstrated in Table~\ref{tab1}, CAFNet, FM-FS, and AMFAM notably outperform Incep-com, HcCNN, and GIIN, highlighting the effectiveness of cross-attention modules, weighted late fusion schemes, and adversarial learning schemes, respectively.
Moreover, our PEMM model attains the highest performance in terms of Avg AUC value (87.6$\%$), surpassing significantly the next three methods (FM-FS: 76.0$\%$, CAFNet: 75.7$\%$, and AMFAM: 75.7$\%$), underscoring the superiority of our approach.
Our PEMM model achieves the top-3 highest values across amolst all categories except PIG-REG, with the highest values in nine categories and the second-highest values in four categories, showing the robustness of our method across eight classification tasks.
In Table~\ref{tab2}, similar phenomena are observable. The proposed PEMM method attains the highest values in five label tasks (PN, BWV, PIG, DaG, and RS) out of eight label tasks, with CAFNet, AMFAM, and FM-SM ranking in the 2nd to 4th positions in terms of Avg ACC.
Specifically, PEMM achieves the highest value of 77.4 $\%$ improve the Avg ACC values of CAFNet (76.8$\%$), AMFAM (76.0$\%$) and FM-FS (75.7$\%$) by 0.7$\%$, 1.3$\%$ and 1.6 $\%$, respectively.

\begin{table*}[]
	\caption{The comparison between our PEMM and currently advanced methods on the SPC dataset in terms of AUC. 
		The highest and second highest values in each column are bolden and italicized respectively. Incep-com: Inception-combined, FM-FS: FusionM4Net-FS, Avg: Averaged ($\%$)}
	\centering
	\tabcolsep=1mm
	\renewcommand\arraystretch{1.2}
	\scalebox{0.6}{
		
		\begin{tabular}{c|cccccccccccccccccc}
			\hline
			\multirow{2}{*}{Methods} & \multicolumn{5}{c}{Diag}                                                                        & \multicolumn{2}{c}{PN}                 & \multicolumn{2}{c}{STR}       & \multicolumn{2}{c}{PIG}       & RS            & \multicolumn{2}{c}{DAG}       & BWV           & \multicolumn{2}{c}{VS}                 & \multirow{2}{*}{\textbf{AVG}} \\ \cline{2-18}
			& BCC           & NEV           & MEL           & MISC                   & SK                     & TYP                    & ATP           & REG           & IR            & REG           & IR            & PRS           & REG           & IR            & PRS           & REG           & IR                     &                               \\ \hline
			Inception-com            & 92.9          & 89.7          & 86.3          & 88.3                   & \textit{91}            & 84.2                   & 79.9          & 87            & 78.9          & 74.9          & 79            & 82.9          & 76.5          & 79.9          & 89.2          & 85.5          & 76.1                   & 83.7                          \\
			HcCNN                    & 94.4          & 87.7          & 85.6          & 88.3                   & 80.4                   & \textit{85.9}          & 78.3          & 87.8          & 77.6          & \textit{83.6} & 81.3          & 81.9          & 77.7          & 82.6          & 89.8          & 87            & 82.7                   & 84.3                          \\
			AMFAM                    & 94.1          & 89.7          & 89.1          & 90.6                   & 81.7                   & 84.5                   & 82.0          & \textit{89.5} & 80.7          & \textbf{85.1} & 83.4          & \textbf{86.7} & 77.7          & 81.9          & 91.1          & \textbf{88.8} & 80.9                   & 85.7                          \\
			FM-FS                    & \textit{95.3} & 92.6          & 89            & \textit{94}            & 89.2                   & \textit{85.9}          & \textit{83.9} & 87.9          & 81.4          & 80.9          & 83.5          & 81.7          & \textit{79.1} & 80.1          & 90.6          & 87.8          & 78                     & 86                            \\
			GIIN                     & 92.8          & 86.8          & 87.6          & 88.8                   & 79.8                   & 80.1                   & \textbf{87.5} & 84.9          & 81.2          & 81.1          & 83.6          & 79            & 78.6          & \textit{83.1} & 90.8          & 80.7          & 75.4                   & 83.6                          \\
			CAFNet                   & \textbf{97.1} & \textit{92.7} & \textbf{92.2} & 92.5                   & \textit{91}            & 81.9                   & 75.3          & 87.4          & \textbf{85.4} & 76.1          & \textit{85}   & \textit{85.4} & 75.2          & 78.7          & \textbf{94.7} & 84.8          & \textit{83.5}          & 85.8                          \\ \hline
			PEMM (Ours)              & 94.7          & \textbf{93.0} & \textit{90.8} & \textit{\textbf{94.9}} & \textit{\textbf{91.7}} & \textit{\textbf{86.7}} & 83.8          & \textbf{90.1} & \textit{84.4} & 79.4          & \textbf{86.1} & 84.9          & \textbf{80.7} & \textbf{84.0} & \textit{93.9} & \textit{88.5} & \textit{\textbf{85.4}} & \textbf{87.6}                 \\ \hline
		\end{tabular}

	}
	\label{tab1}
\end{table*}

\begin{table}[h]
	\caption{The comparison between our PEMM and currently advanced methods on the SPC dataset in terms of accuracy. 
		The highest and second highest values in each column are bolden and italicized respectively. Incep-com: Inception-combined, FM-FS: FusionM4Net-FS, Avg: Averaged ($\%$)}.
	\centering
	\tabcolsep=1mm
	\renewcommand\arraystretch{1.2}
	\scalebox{0.95}{
		
		\begin{tabular}{cccccccccc}
			\hline
			Methods      & PN            & BWV           & VS            & PIG           & STR           & DaG           & RS            & Diag          & AVG           \\ \hline
			Incep-com    & \textit{70.9} & 87.1          & 79.7          & 66.1          & 74.2          & 60.0          & 77.2          & 74.2          & 73.7          \\
			HcCNN        & 70.6          & 87.1          & \textbf{84.8} & 68.6          & 71.6          & \textbf{65.6} & 80.8          & 69.9          & 74.9          \\
			AMFAM        & 70.6          & \textit{88.1} & 83.3          & 70.9          & 74.7          & 63.8          & \textit{82.3} & 75.4          & 76,0          \\
			FM-FS        & \textit{70.9} & 86.8          & 81.8          & \textit{72.4} & 74.4          & 61.0          & \textbf{83.0} & 74.9          & 75.7          \\
			CAFNet       & 70.1          & 87.8          & \textit{84.3} & \textbf{73.4} & \textbf{77.0} & 61.5          & \textit{81.8} & \textbf{78.2} & \textit{76.8} \\
			TFormer      & \textit{70.9} & 86.4          & 83.5          & 68.8          & 74.0          & \textit{64.9} & 81.3          & 73            & 75.3          \\
			PEMM (ours) & \textbf{73.7} & \textbf{88.9} & 82.5          & 71.9 & \textit{76.0} & \textbf{65.6} & \textbf{83.0} & \textit{77.7} & \textbf{77.4} \\ \hline
		\end{tabular}
		
	}
	\label{tab2}
\end{table}

\begin{table}[h]
	\caption{The comprehensive comparison between our AMMFM and other methods in terms of model's parameters.  \textgreater{} : slightly more,  \textgreater{} \textgreater{}: much more }
	\centering
	\tabcolsep=1mm
	\renewcommand\arraystretch{1.2}
	\scalebox{1}{
		\begin{tabular}{cccc}
			\hline
			Method        & Avg AUC ($\%$)       & Avg ACC ($\%$)      & Parameters                        \\ \hline
			Incep-com & 83.7          & 73.7          & \textgreater{}57.4M               \\
			HcCNN         & 84.3          & 74.9          & \textgreater{}65.0M               \\
			AMFAM         & 85.7          & 76            & \textgreater{}51.2M               \\
			FM-FS         & 86            & 75.7          & 54.5M                             \\
			GIIN          & 83.6          & -             & \textgreater{}51.2M               \\
			CAFNet        & 85.8          & 76.8          & \textgreater{}\textgreater{}51.2M \\
			TFormer       & -             & 75.3          & 77.76M                            \\
			PEMM(Ours)    & \textbf{87.6} & \textbf{77.4} & 31.12M                            \\ \hline
		\end{tabular}
	}
	\label{tab5}
	\vspace{-0.1in}
\end{table} 

\begin{table}[h]
	\caption{Further comparison in melanoma-related features ($\%$).}
	\centering
	\tabcolsep=1mm
	\renewcommand\arraystretch{1.2}
	\scalebox{0.7}{		
		\begin{tabular}{ccccccccccc}
			\hline
			\multirow{2}{*}{Metric} & \multirow{2}{*}{Method} & DIAG                     & PN                       & STR                      & PIG                      & RS                       & DaG                      & BWV                      & VS                       & \multirow{2}{*}{Avg}     \\ \cline{3-10}
			&                         & MEL                      & ATP                      & IR                       & IR                       & PRS                      & IR                       & PRS                      & IR                       &                          \\ \hline
			\multirow{7}{*}{AUC}    & Incep-com               & 86.3                     & 79.9                     & 78.9                     & 79                       & 82.9                     & 79.9                     & 89.2                     & 76.1                     & 81.5                     \\
			& HcCNN                   & 85.6                     & 78.3                     & 77.6                     & 81.3                     & 81.9                     & 82.6                     & 89.8                     & 82.7                     & 82.5                     \\
			& AMFAM                   & 89.1                     & 82.0                     & 80.7                     & 83.4                     & 86.7                     & 81.9                     & 91.1                     & 80.9                     & 84.5                     \\
			& FM-FS                   & 89.0                     & 83.9                     & 81.4                     & 83.5                     & 81.7                     & 80.1                     & 90.6                     & 78.9                     & 83.7                     \\
			& GIIN                    & 87.6                     & 87.5                     & 81.2                     & 83.6                     & 79                       & 83.1                     & 90.8                     & 75.4                     & 83.5                     \\
			& CAFNet                  & 92.2                     & 75.3                     & 85.4                     & 85.0                     & 85.4                     & 78.7                     & 94.6                     & 83.4                     & 85.0                     \\
			& PEMM                    & 90.9                     & 83.8                     & 84.4                     & 86.1                     & 84.9                     & 84.0                     & 93.9                     & 85.4                     & \textbf{86.7}            \\ \hline
			\multirow{7}{*}{PRE}    & Incep-com               & 65.3                     & 61.6                     & 52.7                     & 57.8                     & 56.5                     & 70.5                     & 63.0                     & 30.8                     & 57.3                     \\
			& HcCNN                   & 62.8                     & 62.3                     & 52.4                     & 65.1                     & 81.6                     & 69.6                     & 91.9                     & 50.0                     & 67.0                     \\
			& AMFAM                   & 76.2                     & 51.6                     & 54.3                     & 61.3                     & 46.2                     & 82.5                     & 56.0                     & 0.0                      & 53.5                     \\
			& FM-FS                   & 65.7                     & 82.2                     & 56.2                     & 67.6                     & 82.0                     & 67.2                     & 64.9                     & 42.9                     & 68.5                     \\
			& GIIN                    & 65.6                     & 48.4                     & 50.4                     & 82.3                     & 73.5                     & 74.9                     & 67.4                     & 100                      & \textbf{70.3}            \\
			& CAFNet                  & 77.9                     & 50.8                     & 54.8                     & 70.1                     & 76.7                     & 67.8                     & 75.4                     & 58.3                     & 66.5                     \\
			& PEMM                    & 65.4                     & 57.0                     & 52.1                     & 64.5                     & 52.8                     & 78.0                     & 73.3                     & 16.7                     & 57.5                     \\ \hline
			\multirow{7}{*}{SEN}    & Incep-com               & 61.4                     & 48.4                     & 51.1                     & 59.7                     & 66                       & 62.1                     & 77.3                     & 13.3                     & 54.9                     \\
			& HcCNN                   & 58.4                     & 40.9                     & 35.1                     & 55.7                     & 95.2                     & 80.2                     & 92.2                     & 20.0                     & 59.7                     \\
			& AMFAM                   & 65.8                     & 58.5                     & 57.3                     & 67.9                     & 72.1                     & 66.7                     & 75.0                     & 0.0                      & 57.9                     \\
			& FM-FS                   & 62.4                     & 49.5                     & 47.9                     & 58.9                     & 47.1                     & 68.4                     & 66.7                     & 20.0                     & 52.6                     \\
			& GIIN                    & 59.0                     & 77.5                     & 67.0                     & 39.2                     & 21.9                     & 70.1                     & 69.9                     & 3.6                      & 51.0                     \\
			& CAFNet                  & 75.3                     & 65.9                     & 67.1                     & 60.3                     & 42.7                     & 74.1                     & 68.8                     & 45.0                     & 62.4                     \\
			& PEMM                    & 73.3                     & 62.4                     & 57.7                     & 68.4                     & 76.7                     & 71.1                     & 69.6                     & 33.3                     & \textbf{64.1}            \\ \hline
			\multirow{7}{*}{SPE}    & Incep-com               & 88.8                     & 90.7                     & 85.7                     & 80.1                     & 81.3                     & 78.9                     & 89.4                     & 97.5                     & 86.6                     \\
			& HcCNN                   & 88.1                     & 92.4                     & 90.0                     & 86.3                     & 41.5                     & 71.6                     & 65.3                     & 98.4                     & 79.2                     \\
			& AMFAM                   & 91.4                     & 85.6                     & 85.9                     & 83.0                     & 82.6                     & 82.4                     & 90.3                     & 92.4                     & 86.7                     \\
			& FM-FS                   & 88.8                     & 90.1                     & 88.4                     & 88.1                     & 96.2                     & 72.9                     & 91.6                     & 97.8                     & 89.2                     \\
			& GIIN                    & 89.5                     & 79.0                     & 80.3                     & 95.8                     & 96.8                     & 78.8                     & 91.0                     & 100                      & 88.9                     \\
			& CAFNet                  & 93.6                     & 90.2                     & 91.2                     & 89.1                     & 96.5                     & 74.0                     & 95.1                     & 98.7                     & \textbf{91.1}            \\
			& PEMM                    & \multicolumn{1}{l}{88.5} & \multicolumn{1}{l}{87.1} & \multicolumn{1}{l}{85.5} & \multicolumn{1}{l}{84.2} & \multicolumn{1}{l}{84.5} & \multicolumn{1}{l}{80.6} & \multicolumn{1}{l}{93.7} & \multicolumn{1}{l}{93.4} & \multicolumn{1}{l}{87.2} \\ \hline
		\end{tabular}

	}
	\label{tab3}
	\vspace{-0.1in}
\end{table}

For further analysis, we adhere to the methodology outlined in \citep{bi2020multi, fu2022graph} to present the results of melanoma-related features in Table~\ref{tab3}. 
From this table, it is evident that our PEMM model attains the highest performance in terms of Avg AUC at 86.7$\%$ and Avg SEN at 64.1$\%$, thereby affirming the efficacy of our method in detecting melanoma-related features.
Regarding the Avg PRE value, GIIN attains the highest value of 70.3$\%$, surpassing all other methods. We attribute this to the unbalanced distribution of VS-IR (irregular vascular structure), which comprises only 71 positive samples compared to 950 negative samples. This imbalance tends to lead GIIN to over-fit the negative samples of VS-IR, resulting in 100$\%$ values for SPE and PRE but only 3.6$\%$ for SEN. This indicates its effectiveness in detecting negative samples but its limited ability in identifying positive ones.
Conversely, our PEMM achieves the second-highest value (33.3$\%$) in SEN for VS-IR, showcasing its superior performance in detecting positive VS-IR samples even within an extremely unbalanced distribution.

The comparison of model parameters is illustrated in Table~\ref{tab5}. Since there were no descriptions of the parameters for the compared methods in their respective papers, and only the source codes of TFormer and FM-FS are publicly available, we conducted a rough estimation of the parameters for other methods.
Considering that Incep-com, AMFAM, and GIIN do not incorporate an additional third branch and solely utilize two InceptionV3 (57.4Mb) or two ResNet-50 (51.2Mb) as encoders along with fully connected layers as classifiers, we estimate the parameters of Incep-com to be slightly more than 57.4Mb, and the parameters of AMFAM and GIIN to be slightly more than 51.2Mb.
Regarding HcCNN and CAFNet, which incorporate an additional branch, we estimate that their model parameters exceed those of their two encoders (ResNet-50: 51.2Mb).
From the presented table, it is evident that our PEMM model achieves the highest Avg AUC and Avg ACC values while utilizing approximately 60$\%$ fewer parameters compared to the second-best methods, FM-FS (in terms of Avg AUC) and CAFNet (in terms of Avg ACC). This result substantiates the effectiveness of our method.
	
\begin{table}[h]
	\centering
	\caption{Ablation studies of our PEMM in terms of AVG AUC, AVG ACC and model's parameters. FM: Fusion Module, PS: Parameter-Sharing, CA: Cross-Attention, SCA: Shared Cross-Attention, BL: Biased Loss ($\%$). }
	\tabcolsep=1mm
	\renewcommand\arraystretch{1.2}
	\scalebox{1}{
		
		\begin{tabular}{c|ccccccc}
			\hline
			\multicolumn{2}{c}{\multirow{2}{*}{Encoder}} & \multicolumn{2}{c}{FM} & \multirow{2}{*}{BL} & \multirow{2}{*}{AVG AUC} & \multirow{2}{*}{AVG ACC} & \multirow{2}{*}{Parameters} \\ \cline{3-4}
			\multicolumn{2}{c}{}                         & CA        & SCA        &                     &                          &                          &                             \\ \hline
			\multicolumn{2}{c}{Non-PS}                   & \multicolumn{3}{c}{Baseline}                 & 87.1$\pm$0.3                & 76.5$\pm$0.57              & 58.49M                      \\ \hline
			\multicolumn{2}{c}{\multirow{4}{*}{PS}}      &           &            &                     & 86.9$\pm$0.5                & 76.3$\pm$0.43              & 30.14M                      \\
			\multicolumn{2}{c}{}                         & $\checkmark$         &            &                     & 86.7$\pm$0.4                & 75.9$\pm$0.43              & 32.10M                      \\
			\multicolumn{2}{c}{}                         &           & $\checkmark$          &                     & 87.0$\pm$0.1                & 76.6$\pm$0.29              & 31.12M                      \\
			\multicolumn{2}{c}{}                         &           & $\checkmark$          & $\checkmark$                   & 87.1$\pm$0.2                & 76.8$\pm$0.66              & 31.12M                      \\ \hline
		\end{tabular}
		
	}
	\label{tab6}
	\vspace{-0.1in}
\end{table}

\subsection{Ablation studies}
The ablation studies of our PEMM are shown in Table~\ref{tab6} to analyze the effect of three components, i.e., parameter-sharing (PS) encoder, shared cross-attention modules (SCA), biased loss (BL).
The baseline model is a commonly-built multi-modal skin lesion classification model that adopts two individual encoders with a concatenation operation to fuse the features of the final layer of boto modalities, and trained by equally optimization (See Eq.~\ref{eq1}).  
From the data presented in the table, it is apparent that by implementing parameter sharing among encoders, the total parameters of the baseline model experience a significant reduction from 58.49M to 30.14M. Despite this reduction, the decrease in diagnostic performance is negligible, with only a 0.2$\%$ decrease in AVG AUC value and a 0.23$\%$ decrease in AVG ACC (as observed in the 1st and 2nd columns). 
Furthermore, with the incorporation of shared cross-attention (SCA) modules into the PS encoder, there is an improvement in performance from 86.9$\%$ to 87.0$\%$ in AUC value and from 76.3$\%$ to 76.6$\%$ in ACC value, respectively, with only a subtle increase of 0.98M parameters (as shown in the 2nd and 4th columns).
Moreover, the implementation of biased loss (BL) further enhances the performance of the PS-SCA model to 87.1$\%$ in AUC and 76.8$\%$ in ACC values without incurring any increase in computational cost (as shown in the 4th and 5th columns)..
These results illustrate the effectiveness of parameter-sharing networks in parameter reduction, and the efficiency of SCA and BL in enhancing diagnostic accuracy with minimal or no increase in the model's parameters.
In our comparison between CA \citep{he2023co} and our shared CA, we observed that CA does not outperform our SCA and even performs worse than simple concatenation operations (See 2nd-4th columns). This discrepancy may arise from the attention mechanism making the PS network more susceptible to overfitting, thereby resulting in poorer performance compared to concatenations, especially in smaller datasets. However, this issue can be mitigated by employing the PS scheme, akin to the phenomenon observed in \citep{zhang2023tformer}.

\subsection{Other experiments}
\subsubsection{The effect of individual classifiers}
We also investigated the possibility of sharing parameters between the classifiers for the clinical and dermoscopy branches, denoted as $C_C$ and $C_D$, respectively. However, as depicted in Table~\ref{tab8}, compared to the Non-PS classifiers, the PS classifiers exhibit a significant reduction in AUC from 87.1$\%$ to 86.7$\%$ and ACC from 76.8$\%$ to 76.4$\%$. This could be attributed to the sensitivity of fully connected layers to the input.
\begin{table}[h]
	\centering
	\caption{Comparison between our model using parameter-sharing (PS) and non-PS classifiers. ($\%$)}
	\begin{tabular}{c|ccc}
		\hline
		classifiers & \multicolumn{1}{c}{AVG AUC} & AVG ACC   & Parameters(Mb) \\ \hline
		PS         & 86.7$\pm$0.5                   & 76.4$\pm$0.4 & 30.65      \\
		Non-PS     & 87.1$\pm$0.3                   & 76.8$\pm$0.4 & 31.12      \\ \hline
	\end{tabular}
	\label{tab8}
\end{table}
\subsubsection{Comparison of different weight factor $W$}
we conducted an experiment to explore the effect of different weight factors, denoted as $W$, in our biased loss function (Eq.~\ref{eq3}). It is important to note that the weight factor $W$ is assigned to $L_C$, while $0.5 - W$ is allocated to $L_D$.
As illustrated in Table~\ref{tab7}, we observed that the best and second-best overall performances are achieved by setting $W$ to 0.1 and 0.2, respectively, surpassing the method trained using commonly-used equally optimized loss and other settings. This outcome supports our hypothesis that improving classification performance is feasible by leveraging more information from the dermoscopy branch in multi-modal skin lesion classification.
Furthermore, the best overall performance is attained when $W$ is set to 0.1, indicating that specific clinical information can serve as supplementary data to enhance classification performance rather than disregarding it ($W$=0). Conversely, when $W$ is set between 0.3 to 0.5, the corresponding diagnostic performances consistently deteriorate, with the worst performance observed at $W$=0.5. This underscores the significance of incorporating dermoscopy information in the classification process.

\begin{table}[h]
	\centering
	\caption{The effect of different weight factor $W$ in Eq.~\ref{eq3}. EQ: Equally optimization that indicates the model is optimized by the loss function as shwon in Eq.~\ref{eq1}. ($\%$)}
	\begin{tabular}{c|cc}
		\hline
		W   & \multicolumn{1}{c}{AVG AUC}   & AVG ACC              \\ \hline
		0   & \textbf{87.23$\pm$0.36}          & 76.59$\pm$0.51          \\
		0.1 & 87.14$\pm$0.29                   & \textbf{76.84$\pm$0.42} \\
		0.2 & 87.16$\pm$0.41                   & 76.79$\pm$0.13          \\
		0.3 & 87.01$\pm$0.16                   & 76.35$\pm$0.41          \\
		0.4 & 86.56$\pm$0.25                   & 76.12$\pm$0.34          \\
		0.5 & 84.41$\pm$0.18                   & 73.95$\pm$0.38          \\
		EQ  & \multicolumn{1}{c}{87.05$\pm$0.12} & 76.57$\pm$0.29          \\ \hline
	\end{tabular}
	\label{tab7}
\end{table}

\begin{table}[h]
	\caption{Comparisons between single-modal, baseline multi-modal and our PEMM methods based on different backbone. Params: Parameters ($\%$).}
	\centering
	\tabcolsep=1mm
	\renewcommand\arraystretch{1.2}
	\scalebox{1}{
		\begin{tabular}{c|cccc}
			\hline
			Backbone                     & Model    & AVG AUC   & AVG ACC   & Params(Mb)             \\ \hline
			\multirow{4}{*}{ResNet50}    & Derm     & 84.4$\pm$0.3 & 74.2$\pm$0.4 & \multirow{2}{*}{26.68} \\
			& Clic     & 76.8$\pm$0.3 & 67.6$\pm$0.3 &                        \\
			& Baseline & 85.2$\pm$0.1 & 74.3$\pm$0.2 & 55.52                  \\
			& PEMM     & 84.6$\pm$0.2 & 73.3$\pm$0.6 & 36.93                  \\ \hline
			\multirow{4}{*}{DenseNet201} & Derm     & 85.1$\pm$0.2 & 75.1$\pm$0.1 & \multirow{2}{*}{20.17} \\
			& Clic     & 76.8$\pm$0.2 & 68.5$\pm$0.4 &                        \\
			& Baseline & 86.4$\pm$0.4 & 75.7$\pm$0.5 & 44.17                  \\
			& PEMM     & 85.9$\pm$0.3 & 74.6$\pm$0.4 & 29.88                  \\ \hline
			\multirow{4}{*}{Convnext}    & Derm     & 86.5$\pm$0.5 & 75.9$\pm$0.2 & \multirow{2}{*}{29.05} \\
			& Clic     & 77.6$\pm$0.2 & 69.1$\pm$0.2 &                        \\
			& Baseline & 87.0$\pm$0.4 & 76.6$\pm$0.5 & 58.96                  \\
			& PEMM     & 87.1$\pm$0.2 & 76.4$\pm$0.4 & 31.35                  \\ \hline
			\multirow{4}{*}{ST}          & Derm     & 76.3$\pm$0.4 & 86.8$\pm$0.3 & \multirow{2}{*}{28.82} \\
			& Clic     & 69.0$\pm$0.7 & 77.3$\pm$0.4 &                        \\
			& Baseline & 87.1$\pm$0.3 & 76.5$\pm$0.6 & 58.49                  \\
			& PEMM     & 87.1$\pm$0.2 & 76.8$\pm$0.7 & 31.12                  \\ \hline
		\end{tabular}
	}
	\label{tab9}
\end{table}

\begin{table}[h]
	\caption{The effectiveness of our method on the collected ISIC dataset. ($\%$)}
	\centering
	\tabcolsep=1mm
	\renewcommand\arraystretch{1.2}
	\scalebox{1}{
		
		\begin{tabular}{c|cccc}
			\hline
			Backbone                  & Model    & AVG AUC   & AP        & Params(Mb)             \\ \hline
			\multirow{4}{*}{Convnext} & Derm     & 87.3$\pm$0.9 & 87.0$\pm$0.9 & \multirow{2}{*}{29.05} \\
			& Clic     & 83.9$\pm$0.5 & 80.1$\pm$0.4 &                        \\
			& Baseline & 87.2$\pm$0.3 & 84.7$\pm$0.5 & 58.95                  \\
			& PEMM     & 88.3$\pm$0.4 & 86.8$\pm$1.0 & 31.35                  \\ \hline
			\multirow{4}{*}{ST}       & Derm     & 86.6$\pm$0.9 & 85.9$\pm$1.5 & \multirow{2}{*}{28.81} \\
			& Clic     & 85.3$\pm$1.0   & 80.9$\pm$1.6 &                        \\
			& Baseline & 85.0$\pm$2.0 & 84.6$\pm$1.8 & 58.48                  \\
			& PEMM     & 87.2$\pm$1.3 & 86.9$\pm$1.4 & 31.11                  \\ \hline
		\end{tabular}
	}
	\label{tab10}
\end{table}

\subsubsection{The effectiveness our method on different backbones}
In addition to the SwinTransformer (Tiny), we further assessed the effectiveness of our method on different backbones, including ResNet50, DenseNet201, and Convnext (Tiny).
As shown in Table~\ref{tab9}, the proposed PEMM exhibits a decrease in diagnostic performance compared to the Baseline model when utilizing ResNet50 and DenseNet201 as the parameter-sharing encoder. Specifically, for ResNet50 and DenseNet201, both AUC and ACC metrics of PEMM decrease by over 0.5$\%$ compared to the corresponding baseline models. Notably, the AVG ACC value of both backbones even falls below the corresponding single-modality model trained using dermoscopy images (PEMM: 73.3$\%$ vs. Derm: 74.2$\%$ for ResNet50; PEMM: 74.4$\%$ vs. Derm: 75.0$\%$ for DenseNet201).
Conversely, when our PEMM is applied to the Convnext and SwinTransformer backbones, the model's parameters can be significantly compressed (nearly 50$\%$) while maintaining diagnostic accuracy compared to the baseline models, and in the case of SwinTransformer, even performing better.
We believe that this discrepancy may be attributed to differences in the capacity of the backbones. More advanced backbones possess the capacity for parameter-sharing, while traditional backbones may lack this capability.

\subsubsection{The effectiveness our method on the collected ISIC dataset}
To further assess the effectiveness of our PEMM method, we collected a dataset from ISIC Archive for evaluation. Two backbones whose efficacy has been validated on the SPC dataset were continued to be evaluated on this dataset.
As demonstrated in Table~\ref{tab10}, compared to the baseline model, our PEMM achieves better performances while utilizing approximately 50$\%$ fewer model parameters, with both Convnext and SwinTransformer backbones. This underscores that our PEMM consistently achieves efficient multi-modal skin lesion classification across different datasets.
Additionally, it's noteworthy that the multi-modal baseline models perform even worse than the corresponding single-modal models trained solely on dermoscopy images. Conversely, our PEMMs outperform the dermoscopy image-based models in terms of overall performance, indicating the vulnerability of the baseline model when applied to small datasets and the robustness of our PEMM across different datasets.

\section{Conclusion}
In this paper, we introduce a novel Parameter-Efficient Multi-Modal (PEMM) method for skin lesion classification. Our approach offers several key contributions:
Firstly, by sharing the parameters of encoders with strong capacity while retaining individual classifiers, PEMM achieves approximately 50$\%$ compression in model parameters while preserving classification accuracy compared to models employing two separate encoders.
Secondly, our proposed shared cross-attention module enhances modality interactions within the parameter-sharing network (PSN) with fewer parameters compared to commonly-used cross-attention mechanisms.
Finally, we introduce a biased loss function, which leverages the prior knowledge that dermoscopy information is more critical than clinical images. This biased loss guides the PSN to prioritize learning from dermoscopy images, leading to improved optimization and classification.
Extensive experiments validate the effectiveness of our PEMM method in compressing model parameters while maintaining accuracy. 
Furthermore, compared to current state-of-the-art methods, the results demonstrate that PEMM significantly outperforms them while utilizing fewer parameters on the SPC dataset.

	\bibliographystyle{elsarticle-harv} 
	\bibliography{myreference_11_08}

\begin{thebibliography}{37}
\expandafter\ifx\csname natexlab\endcsname\relax\def\natexlab#1{#1}\fi
\providecommand{\url}[1]{\texttt{#1}}
\providecommand{\href}[2]{#2}
\providecommand{\path}[1]{#1}
\providecommand{\DOIprefix}{doi:}
\providecommand{\ArXivprefix}{arXiv:}
\providecommand{\URLprefix}{URL: }
\providecommand{\Pubmedprefix}{pmid:}
\providecommand{\doi}[1]{\href{http://dx.doi.org/#1}{\path{#1}}}
\providecommand{\Pubmed}[1]{\href{pmid:#1}{\path{#1}}}
\providecommand{\bibinfo}[2]{#2}
\ifx\xfnm\relax \def\xfnm[#1]{\unskip,\space#1}\fi
\bibitem[{Aich et~al.(2020)Aich, Yamazaki, Taniguchi and
  Stavness}]{aich2020multi}
\bibinfo{author}{Aich, S.}, \bibinfo{author}{Yamazaki, M.},
  \bibinfo{author}{Taniguchi, Y.}, \bibinfo{author}{Stavness, I.},
  \bibinfo{year}{2020}.
\newblock \bibinfo{title}{Multi-scale weight sharing network for image
  recognition}.
\newblock \bibinfo{journal}{Pattern Recognition Letters} \bibinfo{volume}{131},
  \bibinfo{pages}{348--354}.
\bibitem[{Balch et~al.(2009)Balch, Gershenwald, Soong, Thompson, Atkins, Byrd,
  Buzaid, Cochran, Coit, Ding et~al.}]{balch2009final}
\bibinfo{author}{Balch, C.M.}, \bibinfo{author}{Gershenwald, J.E.},
  \bibinfo{author}{Soong, S.j.}, \bibinfo{author}{Thompson, J.F.},
  \bibinfo{author}{Atkins, M.B.}, \bibinfo{author}{Byrd, D.R.},
  \bibinfo{author}{Buzaid, A.C.}, \bibinfo{author}{Cochran, A.J.},
  \bibinfo{author}{Coit, D.G.}, \bibinfo{author}{Ding, S.}, et~al.,
  \bibinfo{year}{2009}.
\newblock \bibinfo{title}{Final version of 2009 ajcc melanoma staging and
  classification}.
\newblock \bibinfo{journal}{Journal of clinical oncology} \bibinfo{volume}{27},
  \bibinfo{pages}{6199}.
\bibitem[{Bi et~al.(2020)Bi, Feng, Fulham and Kim}]{bi2020multi}
\bibinfo{author}{Bi, L.}, \bibinfo{author}{Feng, D.D.},
  \bibinfo{author}{Fulham, M.}, \bibinfo{author}{Kim, J.},
  \bibinfo{year}{2020}.
\newblock \bibinfo{title}{Multi-label classification of multi-modality skin
  lesion via hyper-connected convolutional neural network}.
\newblock \bibinfo{journal}{Pattern Recognition} \bibinfo{volume}{107},
  \bibinfo{pages}{107502}.
\bibitem[{Dascalu et~al.(2022)Dascalu, Walker, Oron and David}]{dascalu2022non}
\bibinfo{author}{Dascalu, A.}, \bibinfo{author}{Walker, B.},
  \bibinfo{author}{Oron, Y.}, \bibinfo{author}{David, E.},
  \bibinfo{year}{2022}.
\newblock \bibinfo{title}{Non-melanoma skin cancer diagnosis: a comparison
  between dermoscopic and smartphone images by unified visual and sonification
  deep learning algorithms}.
\newblock \bibinfo{journal}{Journal of cancer research and clinical oncology} ,
  \bibinfo{pages}{1--9}.
\bibitem[{Esteva et~al.(2017)Esteva, Kuprel, Novoa, Ko, Swetter, Blau and
  Thrun}]{esteva2017dermatologist}
\bibinfo{author}{Esteva, A.}, \bibinfo{author}{Kuprel, B.},
  \bibinfo{author}{Novoa, R.A.}, \bibinfo{author}{Ko, J.},
  \bibinfo{author}{Swetter, S.M.}, \bibinfo{author}{Blau, H.M.},
  \bibinfo{author}{Thrun, S.}, \bibinfo{year}{2017}.
\newblock \bibinfo{title}{Dermatologist-level classification of skin cancer
  with deep neural networks}.
\newblock \bibinfo{journal}{nature} \bibinfo{volume}{542},
  \bibinfo{pages}{115--118}.
\bibitem[{Fu et~al.(2022)Fu, Bi, Kumar, Fulham and Kim}]{fu2022graph}
\bibinfo{author}{Fu, X.}, \bibinfo{author}{Bi, L.}, \bibinfo{author}{Kumar,
  A.}, \bibinfo{author}{Fulham, M.}, \bibinfo{author}{Kim, J.},
  \bibinfo{year}{2022}.
\newblock \bibinfo{title}{Graph-based intercategory and intermodality network
  for multilabel classification and melanoma diagnosis of skin lesions in
  dermoscopy and clinical images}.
\newblock \bibinfo{journal}{IEEE Transactions on Medical Imaging}
  \bibinfo{volume}{41}, \bibinfo{pages}{3266--3277}.
\bibitem[{Gao et~al.(2024)Gao, He, Meng, Huang, Zhang, Zhang, Xiao and
  Yang}]{gao2024multi}
\bibinfo{author}{Gao, G.}, \bibinfo{author}{He, Y.}, \bibinfo{author}{Meng,
  L.}, \bibinfo{author}{Huang, H.}, \bibinfo{author}{Zhang, D.},
  \bibinfo{author}{Zhang, Y.}, \bibinfo{author}{Xiao, F.},
  \bibinfo{author}{Yang, F.}, \bibinfo{year}{2024}.
\newblock \bibinfo{title}{Multi-view compression and collaboration for skin
  disease diagnosis}.
\newblock \bibinfo{journal}{Expert Systems with Applications} ,
  \bibinfo{pages}{123395}.
\bibitem[{Ge et~al.(2017)Ge, Demyanov, Chakravorty, Bowling and
  Garnavi}]{ge2017skin}
\bibinfo{author}{Ge, Z.}, \bibinfo{author}{Demyanov, S.},
  \bibinfo{author}{Chakravorty, R.}, \bibinfo{author}{Bowling, A.},
  \bibinfo{author}{Garnavi, R.}, \bibinfo{year}{2017}.
\newblock \bibinfo{title}{Skin disease recognition using deep saliency features
  and multimodal learning of dermoscopy and clinical images}, in:
  \bibinfo{booktitle}{Medical Image Computing and Computer Assisted
  Intervention- MICCAI 2017: 20th International Conference, Quebec City, QC,
  Canada, September 11-13, 2017, Proceedings, Part III 20},
  \bibinfo{organization}{Springer}. pp. \bibinfo{pages}{250--258}.
\bibitem[{Gu et~al.(2019)Gu, Ge, Bonnington and Zhou}]{gu2019progressive}
\bibinfo{author}{Gu, Y.}, \bibinfo{author}{Ge, Z.},
  \bibinfo{author}{Bonnington, C.P.}, \bibinfo{author}{Zhou, J.},
  \bibinfo{year}{2019}.
\newblock \bibinfo{title}{Progressive transfer learning and adversarial domain
  adaptation for cross-domain skin disease classification}.
\newblock \bibinfo{journal}{IEEE journal of biomedical and health informatics}
  \bibinfo{volume}{24}, \bibinfo{pages}{1379--1393}.
\bibitem[{He et~al.(2016)He, Zhang, Ren and Sun}]{he2016deep}
\bibinfo{author}{He, K.}, \bibinfo{author}{Zhang, X.}, \bibinfo{author}{Ren,
  S.}, \bibinfo{author}{Sun, J.}, \bibinfo{year}{2016}.
\newblock \bibinfo{title}{Deep residual learning for image recognition}, in:
  \bibinfo{booktitle}{Proceedings of the IEEE conference on computer vision and
  pattern recognition}, pp. \bibinfo{pages}{770--778}.
\bibitem[{He et~al.(2023)He, Wang, Zhao and Chen}]{he2023co}
\bibinfo{author}{He, X.}, \bibinfo{author}{Wang, Y.}, \bibinfo{author}{Zhao,
  S.}, \bibinfo{author}{Chen, X.}, \bibinfo{year}{2023}.
\newblock \bibinfo{title}{Co-attention fusion network for multimodal skin
  cancer diagnosis}.
\newblock \bibinfo{journal}{Pattern Recognition} \bibinfo{volume}{133},
  \bibinfo{pages}{108990}.
\bibitem[{Huang et~al.(2017)Huang, Liu, Van Der~Maaten and
  Weinberger}]{huang2017densely}
\bibinfo{author}{Huang, G.}, \bibinfo{author}{Liu, Z.}, \bibinfo{author}{Van
  Der~Maaten, L.}, \bibinfo{author}{Weinberger, K.Q.}, \bibinfo{year}{2017}.
\newblock \bibinfo{title}{Densely connected convolutional networks}, in:
  \bibinfo{booktitle}{Proceedings of the IEEE conference on computer vision and
  pattern recognition}, pp. \bibinfo{pages}{4700--4708}.
\bibitem[{Huang et~al.(2022)Huang, You, Zheng, Wang, Qian and
  Yamasaki}]{huang2022learning}
\bibinfo{author}{Huang, L.}, \bibinfo{author}{You, S.}, \bibinfo{author}{Zheng,
  M.}, \bibinfo{author}{Wang, F.}, \bibinfo{author}{Qian, C.},
  \bibinfo{author}{Yamasaki, T.}, \bibinfo{year}{2022}.
\newblock \bibinfo{title}{Learning where to learn in cross-view self-supervised
  learning}, in: \bibinfo{booktitle}{Proceedings of the IEEE/CVF Conference on
  Computer Vision and Pattern Recognition}, pp. \bibinfo{pages}{14451--14460}.
\bibitem[{Izmailov et~al.(2018)Izmailov, Podoprikhin, Garipov, Vetrov and
  Wilson}]{izmailov2018averaging}
\bibinfo{author}{Izmailov, P.}, \bibinfo{author}{Podoprikhin, D.},
  \bibinfo{author}{Garipov, T.}, \bibinfo{author}{Vetrov, D.},
  \bibinfo{author}{Wilson, A.G.}, \bibinfo{year}{2018}.
\newblock \bibinfo{title}{Averaging weights leads to wider optima and better
  generalization}.
\newblock \bibinfo{journal}{arXiv preprint arXiv:1803.05407} .
\bibitem[{Kawahara et~al.(2018)Kawahara, Daneshvar, Argenziano and
  Hamarneh}]{kawahara2018seven}
\bibinfo{author}{Kawahara, J.}, \bibinfo{author}{Daneshvar, S.},
  \bibinfo{author}{Argenziano, G.}, \bibinfo{author}{Hamarneh, G.},
  \bibinfo{year}{2018}.
\newblock \bibinfo{title}{Seven-point checklist and skin lesion classification
  using multitask multimodal neural nets}.
\newblock \bibinfo{journal}{IEEE journal of biomedical and health informatics}
  \bibinfo{volume}{23}, \bibinfo{pages}{538--546}.
\bibitem[{Kingma and Ba(2014)}]{kingma2014adam}
\bibinfo{author}{Kingma, D.P.}, \bibinfo{author}{Ba, J.}, \bibinfo{year}{2014}.
\newblock \bibinfo{title}{Adam: A method for stochastic optimization}.
\newblock \bibinfo{journal}{arXiv preprint arXiv:1412.6980} .
\bibitem[{Kolarsick et~al.(2011)Kolarsick, Kolarsick and
  Goodwin}]{kolarsick2011anatomy}
\bibinfo{author}{Kolarsick, P.A.}, \bibinfo{author}{Kolarsick, M.A.},
  \bibinfo{author}{Goodwin, C.}, \bibinfo{year}{2011}.
\newblock \bibinfo{title}{Anatomy and physiology of the skin}.
\newblock \bibinfo{journal}{Journal of the Dermatology Nurses' Association}
  \bibinfo{volume}{3}, \bibinfo{pages}{203--213}.
\bibitem[{Liu et~al.(2021)Liu, Lin, Cao, Hu, Wei, Zhang, Lin and
  Guo}]{liu2021swin}
\bibinfo{author}{Liu, Z.}, \bibinfo{author}{Lin, Y.}, \bibinfo{author}{Cao,
  Y.}, \bibinfo{author}{Hu, H.}, \bibinfo{author}{Wei, Y.},
  \bibinfo{author}{Zhang, Z.}, \bibinfo{author}{Lin, S.}, \bibinfo{author}{Guo,
  B.}, \bibinfo{year}{2021}.
\newblock \bibinfo{title}{Swin transformer: Hierarchical vision transformer
  using shifted windows}, in: \bibinfo{booktitle}{Proceedings of the IEEE/CVF
  international conference on computer vision}, pp.
  \bibinfo{pages}{10012--10022}.
\bibitem[{Liu et~al.(2022a)Liu, Mao, Wu, Feichtenhofer, Darrell and
  Xie}]{liu2022convnet}
\bibinfo{author}{Liu, Z.}, \bibinfo{author}{Mao, H.}, \bibinfo{author}{Wu,
  C.Y.}, \bibinfo{author}{Feichtenhofer, C.}, \bibinfo{author}{Darrell, T.},
  \bibinfo{author}{Xie, S.}, \bibinfo{year}{2022}a.
\newblock \bibinfo{title}{A convnet for the 2020s}, in:
  \bibinfo{booktitle}{Proceedings of the IEEE/CVF conference on computer vision
  and pattern recognition}, pp. \bibinfo{pages}{11976--11986}.
\bibitem[{Liu et~al.(2022b)Liu, Xiong and Jiang}]{liu2022ci}
\bibinfo{author}{Liu, Z.}, \bibinfo{author}{Xiong, R.}, \bibinfo{author}{Jiang,
  T.}, \bibinfo{year}{2022}b.
\newblock \bibinfo{title}{Ci-net: clinical-inspired network for automated skin
  lesion recognition}.
\newblock \bibinfo{journal}{IEEE Transactions on Medical Imaging}
  \bibinfo{volume}{42}, \bibinfo{pages}{619--632}.
\bibitem[{Pacheco et~al.(2020)Pacheco, Lima, Salomao, Krohling, Biral,
  de~Angelo, Alves~Jr, Esgario, Simora, Castro et~al.}]{pacheco2020pad}
\bibinfo{author}{Pacheco, A.G.}, \bibinfo{author}{Lima, G.R.},
  \bibinfo{author}{Salomao, A.S.}, \bibinfo{author}{Krohling, B.},
  \bibinfo{author}{Biral, I.P.}, \bibinfo{author}{de~Angelo, G.G.},
  \bibinfo{author}{Alves~Jr, F.C.}, \bibinfo{author}{Esgario, J.G.},
  \bibinfo{author}{Simora, A.C.}, \bibinfo{author}{Castro, P.B.}, et~al.,
  \bibinfo{year}{2020}.
\newblock \bibinfo{title}{Pad-ufes-20: A skin lesion dataset composed of
  patient data and clinical images collected from smartphones}.
\newblock \bibinfo{journal}{Data in brief} \bibinfo{volume}{32},
  \bibinfo{pages}{106221}.
\bibitem[{Ricci~Lara et~al.(2023)Ricci~Lara, Rodr{\'\i}guez~Kowalczuk,
  Lisa~Eliceche, Ferraresso, Luna, Benitez and Mazzuoccolo}]{ricci2023dataset}
\bibinfo{author}{Ricci~Lara, M.A.}, \bibinfo{author}{Rodr{\'\i}guez~Kowalczuk,
  M.V.}, \bibinfo{author}{Lisa~Eliceche, M.}, \bibinfo{author}{Ferraresso,
  M.G.}, \bibinfo{author}{Luna, D.R.}, \bibinfo{author}{Benitez, S.E.},
  \bibinfo{author}{Mazzuoccolo, L.D.}, \bibinfo{year}{2023}.
\newblock \bibinfo{title}{A dataset of skin lesion images collected in
  argentina for the evaluation of ai tools in this population}.
\newblock \bibinfo{journal}{Scientific Data} \bibinfo{volume}{10},
  \bibinfo{pages}{712}.
\bibitem[{Sch{\"u}rholt et~al.(2021)Sch{\"u}rholt, Kostadinov and
  Borth}]{schurholt2021self}
\bibinfo{author}{Sch{\"u}rholt, K.}, \bibinfo{author}{Kostadinov, D.},
  \bibinfo{author}{Borth, D.}, \bibinfo{year}{2021}.
\newblock \bibinfo{title}{Self-supervised representation learning on neural
  network weights for model characteristic prediction}.
\newblock \bibinfo{journal}{Advances in Neural Information Processing Systems}
  \bibinfo{volume}{34}, \bibinfo{pages}{16481--16493}.
\bibitem[{Siegel et~al.(2022)Siegel, Miller, Fuchs and
  Jemal}]{siegel2022cancer}
\bibinfo{author}{Siegel, R.L.}, \bibinfo{author}{Miller, K.D.},
  \bibinfo{author}{Fuchs, H.E.}, \bibinfo{author}{Jemal, A.},
  \bibinfo{year}{2022}.
\newblock \bibinfo{title}{Cancer statistics, 2022}.
\newblock \bibinfo{journal}{CA: a cancer journal for clinicians}
  \bibinfo{volume}{72}, \bibinfo{pages}{7--33}.
\bibitem[{Sun et~al.(2016)Sun, Yang, Sun and Wang}]{sun2016benchmark}
\bibinfo{author}{Sun, X.}, \bibinfo{author}{Yang, J.}, \bibinfo{author}{Sun,
  M.}, \bibinfo{author}{Wang, K.}, \bibinfo{year}{2016}.
\newblock \bibinfo{title}{A benchmark for automatic visual classification of
  clinical skin disease images}, in: \bibinfo{booktitle}{Computer Vision--ECCV
  2016: 14th European Conference, Amsterdam, The Netherlands, October 11-14,
  2016, Proceedings, Part VI 14}, \bibinfo{organization}{Springer}. pp.
  \bibinfo{pages}{206--222}.
\bibitem[{Tang et~al.(2020)Tang, Liang, Yan, Xiang and Zhang}]{tang2020gp}
\bibinfo{author}{Tang, P.}, \bibinfo{author}{Liang, Q.}, \bibinfo{author}{Yan,
  X.}, \bibinfo{author}{Xiang, S.}, \bibinfo{author}{Zhang, D.},
  \bibinfo{year}{2020}.
\newblock \bibinfo{title}{Gp-cnn-dtel: Global-part cnn model with
  data-transformed ensemble learning for skin lesion classification}.
\newblock \bibinfo{journal}{IEEE journal of biomedical and health informatics}
  \bibinfo{volume}{24}, \bibinfo{pages}{2870--2882}.
\bibitem[{Tang et~al.(2022)Tang, Yan, Nan, Xiang, Krammer and
  Lasser}]{tang2022fusionm4net}
\bibinfo{author}{Tang, P.}, \bibinfo{author}{Yan, X.}, \bibinfo{author}{Nan,
  Y.}, \bibinfo{author}{Xiang, S.}, \bibinfo{author}{Krammer, S.},
  \bibinfo{author}{Lasser, T.}, \bibinfo{year}{2022}.
\newblock \bibinfo{title}{Fusionm4net: A multi-stage multi-modal learning
  algorithm for multi-label skin lesion classification}.
\newblock \bibinfo{journal}{Medical Image Analysis} \bibinfo{volume}{76},
  \bibinfo{pages}{102307}.
\bibitem[{Tao et~al.(2022)Tao, Wang, Zhu, Dong, Song, Huang and
  Dai}]{tao2022exploring}
\bibinfo{author}{Tao, C.}, \bibinfo{author}{Wang, H.}, \bibinfo{author}{Zhu,
  X.}, \bibinfo{author}{Dong, J.}, \bibinfo{author}{Song, S.},
  \bibinfo{author}{Huang, G.}, \bibinfo{author}{Dai, J.}, \bibinfo{year}{2022}.
\newblock \bibinfo{title}{Exploring the equivalence of siamese self-supervised
  learning via a unified gradient framework}, in:
  \bibinfo{booktitle}{Proceedings of the IEEE/CVF Conference on Computer Vision
  and Pattern Recognition}, pp. \bibinfo{pages}{14431--14440}.
\bibitem[{Togawa et~al.(2023)Togawa, Yamamoto and
  Matsue}]{togawa2023comparison}
\bibinfo{author}{Togawa, Y.}, \bibinfo{author}{Yamamoto, Y.},
  \bibinfo{author}{Matsue, H.}, \bibinfo{year}{2023}.
\newblock \bibinfo{title}{Comparison of images obtained using four dermoscope
  imaging devices: An observational study}.
\newblock \bibinfo{journal}{JEADV Clinical Practice} \bibinfo{volume}{2},
  \bibinfo{pages}{888--892}.
\bibitem[{Wang et~al.(2023)Wang, Xu, Chen, Tong, Chen, Hu and
  Lin}]{wang2023adaptive}
\bibinfo{author}{Wang, H.}, \bibinfo{author}{Xu, Y.}, \bibinfo{author}{Chen,
  Q.}, \bibinfo{author}{Tong, R.}, \bibinfo{author}{Chen, Y.W.},
  \bibinfo{author}{Hu, H.}, \bibinfo{author}{Lin, L.}, \bibinfo{year}{2023}.
\newblock \bibinfo{title}{Adaptive decomposition and shared weight volumetric
  transformer blocks for efficient patch-free 3d medical image segmentation}.
\newblock \bibinfo{journal}{IEEE Journal of Biomedical and Health Informatics}
  .
\bibitem[{Wang et~al.(2022)Wang, Feng, Zhang, Zhou, Liu, Goh and
  Zhen}]{wang2022adversarial}
\bibinfo{author}{Wang, Y.}, \bibinfo{author}{Feng, Y.}, \bibinfo{author}{Zhang,
  L.}, \bibinfo{author}{Zhou, J.T.}, \bibinfo{author}{Liu, Y.},
  \bibinfo{author}{Goh, R.S.M.}, \bibinfo{author}{Zhen, L.},
  \bibinfo{year}{2022}.
\newblock \bibinfo{title}{Adversarial multimodal fusion with attention
  mechanism for skin lesion classification using clinical and dermoscopic
  images}.
\newblock \bibinfo{journal}{Medical Image Analysis} \bibinfo{volume}{81},
  \bibinfo{pages}{102535}.
\bibitem[{Wang et~al.(2020)Wang, Sun, Lu and Yao}]{wang2020learning}
\bibinfo{author}{Wang, Y.}, \bibinfo{author}{Sun, F.}, \bibinfo{author}{Lu,
  M.}, \bibinfo{author}{Yao, A.}, \bibinfo{year}{2020}.
\newblock \bibinfo{title}{Learning deep multimodal feature representation with
  asymmetric multi-layer fusion}, in: \bibinfo{booktitle}{Proceedings of the
  28th ACM International Conference on Multimedia}, pp.
  \bibinfo{pages}{3902--3910}.
\bibitem[{Yang et~al.(2018)Yang, Sun, Liang and Rosin}]{yang2018clinical}
\bibinfo{author}{Yang, J.}, \bibinfo{author}{Sun, X.}, \bibinfo{author}{Liang,
  J.}, \bibinfo{author}{Rosin, P.L.}, \bibinfo{year}{2018}.
\newblock \bibinfo{title}{Clinical skin lesion diagnosis using representations
  inspired by dermatologist criteria}, in: \bibinfo{booktitle}{Proceedings of
  the IEEE Conference on Computer Vision and Pattern Recognition}, pp.
  \bibinfo{pages}{1258--1266}.
\bibitem[{Yang et~al.(2019)Yang, Wu, Liang, Sun, Cheng, Rosin and
  Wang}]{yang2019self}
\bibinfo{author}{Yang, J.}, \bibinfo{author}{Wu, X.}, \bibinfo{author}{Liang,
  J.}, \bibinfo{author}{Sun, X.}, \bibinfo{author}{Cheng, M.M.},
  \bibinfo{author}{Rosin, P.L.}, \bibinfo{author}{Wang, L.},
  \bibinfo{year}{2019}.
\newblock \bibinfo{title}{Self-paced balance learning for clinical skin disease
  recognition}.
\newblock \bibinfo{journal}{IEEE transactions on neural networks and learning
  systems} \bibinfo{volume}{31}, \bibinfo{pages}{2832--2846}.
\bibitem[{Yao et~al.(2021)Yao, Shen, Xu, Liu, Zhang, Xing, Shao, Kaffenberger
  and Xu}]{yao2021single}
\bibinfo{author}{Yao, P.}, \bibinfo{author}{Shen, S.}, \bibinfo{author}{Xu,
  M.}, \bibinfo{author}{Liu, P.}, \bibinfo{author}{Zhang, F.},
  \bibinfo{author}{Xing, J.}, \bibinfo{author}{Shao, P.},
  \bibinfo{author}{Kaffenberger, B.}, \bibinfo{author}{Xu, R.X.},
  \bibinfo{year}{2021}.
\newblock \bibinfo{title}{Single model deep learning on imbalanced small
  datasets for skin lesion classification}.
\newblock \bibinfo{journal}{IEEE transactions on medical imaging}
  \bibinfo{volume}{41}, \bibinfo{pages}{1242--1254}.
\bibitem[{Yap et~al.(2018)Yap, Yolland and Tschandl}]{yap2018multimodal}
\bibinfo{author}{Yap, J.}, \bibinfo{author}{Yolland, W.},
  \bibinfo{author}{Tschandl, P.}, \bibinfo{year}{2018}.
\newblock \bibinfo{title}{Multimodal skin lesion classification using deep
  learning}.
\newblock \bibinfo{journal}{Experimental dermatology} \bibinfo{volume}{27},
  \bibinfo{pages}{1261--1267}.
\bibitem[{Zhang et~al.(2023)Zhang, Xie and Chen}]{zhang2023tformer}
\bibinfo{author}{Zhang, Y.}, \bibinfo{author}{Xie, F.}, \bibinfo{author}{Chen,
  J.}, \bibinfo{year}{2023}.
\newblock \bibinfo{title}{Tformer: A throughout fusion transformer for
  multi-modal skin lesion diagnosis}.
\newblock \bibinfo{journal}{Computers in Biology and Medicine}
  \bibinfo{volume}{157}, \bibinfo{pages}{106712}.

\end{thebibliography}
\end{document}